# Low-threshold lasing of optically pumped micropillar lasers with Al$_{0.2}$Ga$_{0.8}$As/Al$_{0.9}$Ga$_{0.1}$As distributed Bragg reflectors

Ching-Wen Shih,[1] Imad Limame,[1] Sebastian Krüger,[1] Chirag C. Palekar,[1] Aris Koulas-Simos,[1] Daniel Brunner,[2] Stephan Reitzenstein[1, a)] [1]*Institut für Festkörperphysik, Technische Universität Berlin, Hardenbergstraße 36, D-10623 Berlin, Germany*

[2]*Département d'Optique P. M. Duffieux, Institut FEMTO-ST, Université Franche-Comté CNRS UMR 6174, Besançon, France*

We report on the design, realization and characterization of optically pumped micropillar lasers with low-absorbing Al$_{0.2}$Ga$_{0.8}$As/Al$_{0.9}$Ga$_{0.1}$As dielectric Bragg reflectors (DBRs) instead of commonly used GaAs/AlGaAs DBRs. A layer of (In, Ga)As quantum dots (QDs) is embedded in the GaAs $\lambda$-cavity of as an active medium. We experimentally study the lasing characteristics of the fabricated micropillars by means of low-temperature photoluminescence with varying pump laser's wavelength between 532 nm and 899 nm. The incorporation of 20% Al content in the DBRs opens an optical pumping window from 700 nm to 820 nm, where the excitation laser light can effectively reach the GaAs cavity above its bandgap, while remaining transparent to the DBRs. This results in a substantially improved pump efficiency, a low lasing threshold, and a high thermal stability. Pump laser wavelengths outside of the engineered spectral window lead to low pump efficiency due to strong absorption by the top DBR, or inefficient excitation of pump-level excitons, respectively. The superiority of the absorption-free modified DBRs is demonstrated by simply switching the pump laser wavelength from 671 nm to 708 nm, which crosses the DBRs absorption edge and drastically reduces the lasing threshold by more than an order of magnitude from $(363.5 \pm 18.5)\,\mu$W to $(12.8 \pm 0.3)\,\mu$W.

Semiconductor quantum dots (QDs) embedded in an optical microcavity have become an attractive and powerful platform for the study of cavity quantum electrodynamics (cQED) in the solid state, but also to realize bright photonic sources to enable advanced technologies, including quantum communication,[1,2] optical sensing,[3] and photonic neuromorphic computation.[4,5] An interesting experimental platform is the quantum dot micropillar system, where the quantum dots are located in a central cavity layer, sandwiched between epitaxially grown dielectric Bragg mirrors (DBRs) to offer a high Q-factor, a small mode volume,

a) stephan.reitzenstein@physik.tu-berlin.de

and therefore an enhanced Purcell factor.[6,7] Particularly, partial injection locking has recently been reported with micropillar lasers,[8] making them a promising alternative to conventional vertical cavity surface emitting lasers (VCSELs) as photonic neurons with cQED-enhanced properties and a so far unparalleled integration density. In particular for neuromorphic computing applications, spectral homogeneity across the array is the key and has been recently addressed by means of tuning the diameter of each micropillar individually, leading to a notable progress in reducing the inhomogeneous broadening of fabricated micropillar arrays.[5,9]

Furthermore, the number of neurons, i.e. microlasers, in this case is of fundamental importance. In order to increase the number of photonic neurons in an energy-efficient manner, it is crucial to maximize their power efficiency and reduce their lasing threshold. Although electrically injected micropillar lasers have a low-lasing threshold,[10] upscaling dense arrays of such devices requires a complex electrical contact layout and even poses limitations on lasers' rf-modulation bandwidth. Optically pumped micropillar lasers based on GaAs material systems, on the other hand, are often reported to suffer from a low power conversion efficiency (PCE) around 3% due to the strong optical pump power absorption in the top Al(Ga)As/GaAs DBRs before it reaches the GaAs cavity where the quantum dots are located.[11] As a result, despite cavity-enhanced operation, the high lasing threshold of optically pumped micropillar lasers hinders the upscaling of photonic neural networks leveraging such devices as neurons.

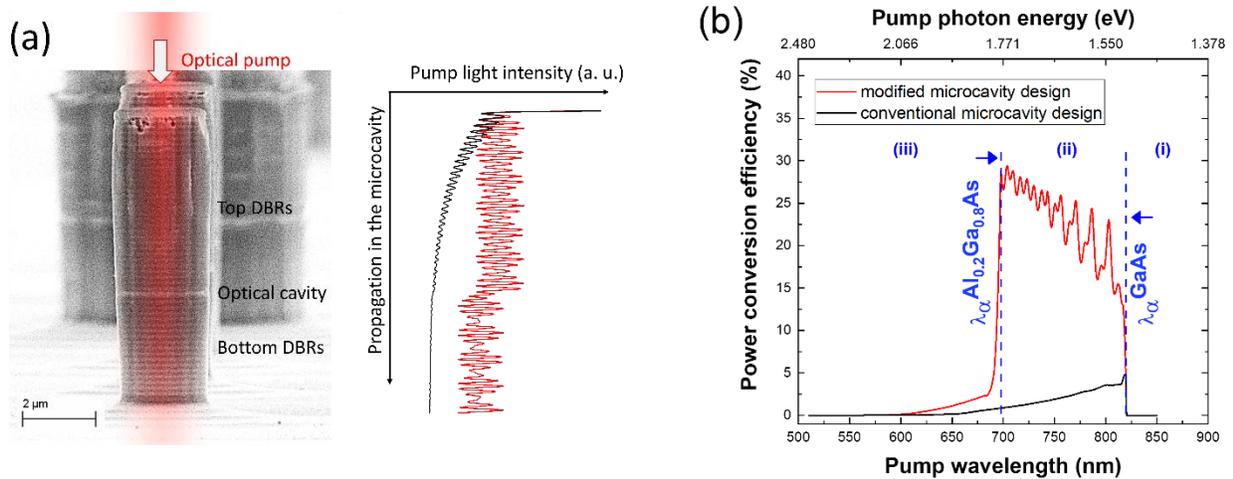

FIG. 1. (a) Scanning electron microscope (SEM) image (side view) of a fabricated micropillar lasers along with the pump light intensity propagation in such microcavity. Black: the pump light decays exponentially when the DBRs is absorbing. Red: the pump light intensity remains unabsorbed until it reaches the optical cavity. (b) Simulated PCE as a function of the pump wavelength with both modified and conventional microcavity design. Regime (i), (ii), and (iii) discussed in the text are marked.

A simple and yet long-overlooked solution to the low-PCE problem has been proposed recently, namely to replace the conventional Al(Ga)As/GaAs DBRs with Al(Ga)As/AlGaAs DBRs to displace the absorption edge of the Al(Ga)As DBRs relative to the GaAs cavity, opening an ideal optical pumping window over a wide wavelength range.[11,12] Figure 1(a) illustrates the propagation of pump light in a micropillar, showing for the conventional design the exponential damping of pump light due to absorption in the top DBR before reaching the optical cavity (black data), in contrast to the low-absorbing design where pump light reaches the optical cavity without such absorption (red data). The simple modification is fully compatible with the conventional fabrication environment, but up to now is not commonly adopted by the micropillar community. In this work, we report on the realization and systematical study of such modified micropillar lasers with the $Al_{0.2}Ga_{0.8}As/Al_{0.9}Ga_{0.1}As$ DBRs and stress its role to the reduction of lasing threshold.

The epitaxial layers of our planar micropillar samples are grown by means of metal-organic chemical vapor deposition (MOCVD). The resonator consists of 33.5 pairs of $\lambda/4n$-thick $Al_{0.9}Ga_{0.1}As/Al_{0.2}Ga_{0.8}As$ layers in the bottom DBR, followed by a single layer of self-assembled InGaAs QDs embedded in a $\lambda/n$-thick GaAs cavity, and 29.5 pairs of $\lambda/4n$-thick $Al_{0.9}Ga_{0.1}As/Al_{0.2}Ga_{0.8}As$ forming the top DBR, which is capped with a $\lambda/4n$-thick GaAs layer to prevent oxidation. The InGaAs QDs are optimized to have a high density of $\sim 1.4 \times 10^{10}$ cm$^{-2}$ providing sufficient optical gain for lasing. Following the growth, micropillar lasers are processed by means of electron beam lithography (EBL), the details of which have been reported in our previous work.[5]

The power conversion efficiency is defined as the percentage of the pump power that is absorbed within the GaAs cavity, from where the photons are converted into excitons in the QDs which contribute to lasing. Since the PCE linearly scales with the applied pump power that is not absorbed or reflected by the DBRs, its value strongly influences the transparency threshold and the lasing threshold. As a first step, this quantity is determined via the transfer matrix method (TMM). In the calculation, absorption from the wetting layer and the QDs are omitted since the active layer thickness are too thin to effectively absorb the pump laser. The optical constants of the (Al)GaAs at cryogenic temperature are estimated by extrapolation and interpolation from reported values.[13–15] Figure 1(b) shows the PCE of our modified design compared to the conventional design with 24/27 pairs $Al_{0.9}Ga_{0.1}As/GaAs$ DBRs.[11] Note that the conventional DBR design has 6 mirror pairs less than the modified design in order to provide a similar Q-factor and Purcell factor which is discussed in the supplementary material.

Separated by the optical absorption edge of GaAs ($\lambda_\alpha^{GaAs}$) around 820 nm and the optical absorption edge of $Al_{0.2}Ga_{0.9}As$ ($\lambda_\alpha^{Al_{0.2}Ga_{0.8}As}$) around 700 nm, three optical pumping schemes can be readily identified: (i) When the pump wavelength is longer than $\lambda_\alpha^{GaAs}$, there is only negligible absorption occurring in the InGaAs wetting layer and QDs, leading to low PCE. (ii) When the pump wavelength is shorter than $\lambda_\alpha^{GaAs}$ but longer than $\lambda_\alpha^{Al_{0.2}Ga_{0.8}As}$, the modified DBRs design becomes superior to the conventional DBRs, because the modified top DBR is transparent at the laser wavelength such that the pump light can efficiently reach the GaAs cavity layer where it is then absorbed, creating excitons. One can also notice that the PCE of the modified DBR design rises with shortening pump wavelength thanks to the increasing optical absorption in the GaAs cavity, while the PCE of the conventional design decreases almost monotonously due to the strong depletion of the pump power. (iii) When the pump wavelength is shorter than $\lambda_\alpha^{Al_{0.2}Ga_{0.8}As}$, even the $Al_{0.2}Ga_{0.8}As$ layers in the modified DBRs start absorbing, which drastically decreases the PCE. Note that absorption is not the only factor that could affect the PCE, reflectance fringes from the DBRs sideband but also plays an important role and leads to the oscillatory behavior of PCE.

Scheme (ii) clearly provides the better pumping condition, and the PCE of the modified design can potentially reach 29%, which is nearly 6 times higher than the 5% of the conventional design. The field propagation in both designs at commonly used pump wavelengths, such as 532 nm, 671 nm, and 781 nm, is visualized in Fig. 2. In the conventional design presented in Fig. 2(a), the light field is strongly absorbed in the DBR layers and decays exponentially regardless of the pump laser wavelength. In the modified design shown in Fig. 2(b), the DBRs becomes transparent to the 781 nm pump laser and the optical power is only absorbed within the GaAs cavity. Note that despite the hugely improved PCE, more than 70% of the pump power is either reflected or still transmitted to the bottom GaAs wafer due to the finite thickness of the GaAs λ-cavity; the details of which are shown in the supplementary material.

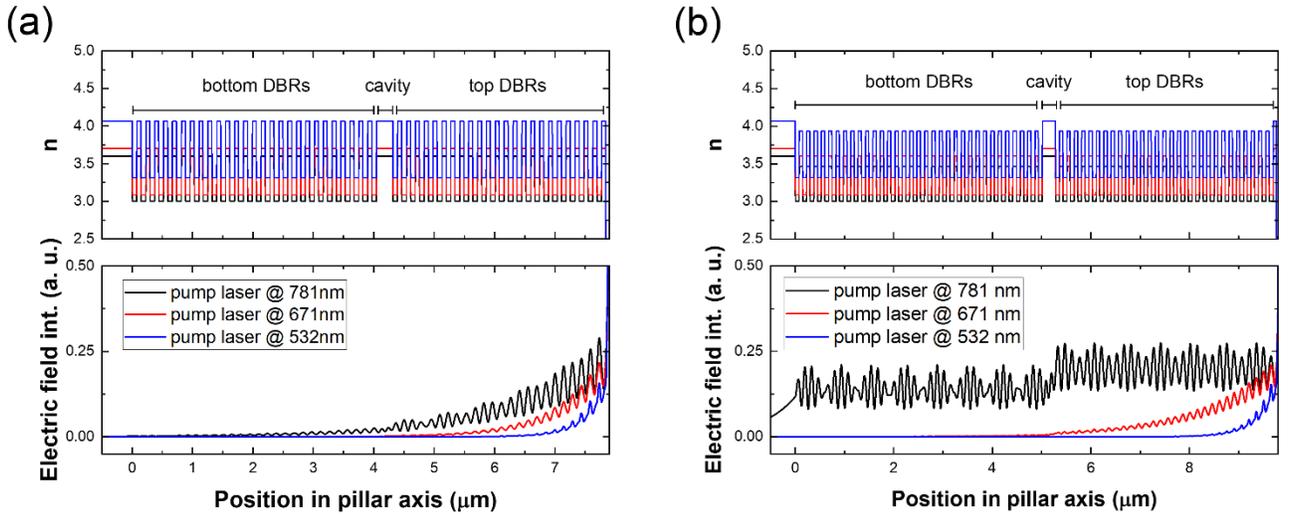

FIG. 2. (upper panels) Refractive index and (lower panels) normalized electric field intensity under different pump laser wavelengths. Microcavity based on (a) conventional design with 24/27 pairs of DBRs, and (b) modified design same as our experimental design. The pump laser is incident from the right onto the microcavity, with the GaAs (001) substrate located on the left.

In order to experimentally support the numerical results, pump-wavelength-dependent I/O measurements are carried out on our fabricated micropillar lasers in a typical micro-photoluminescence setup with a He-flow cryostat operated at 77 K. For the optical pumping, a 532 nm, a 671 nm and a 781 nm diode pump solid state laser (DPSSL) along with a Ti:sapphire tunable continuous-wave laser (Ti-Sa) are integrated into the setup to investigate the three above mentioned pumping schemes (i, ii, and iii). A variable magnification beam expander is installed in front of the lasers to carefully adjust the laser beamwidth, leading to a focused spot size around 1-2 µm with an optical objective of NA = 0.4 in front of the sample holder. In the detection path, a linear polarizer followed by a half waveplate are installed in front of the input slit of a high-resolution spectrometer to distinguish between the two linearly polarized modes of fundamental mode, which arises from unintentional micropillar ellipticity in fabrication.[16,17] In the following, they are labeled as mode 1 and mode 2, respectively.

Figure 3 shows the optical characteristics of a micropillar laser with a diameter of 3.6 µm pumped by a 671 and a 781 nm laser, corresponding to the absorbing scheme (iii) and the low-absorbing scheme (ii), respectively. Figures 3(a) and 3(d) depict the normalized emission spectra of mode 2 at different pump power. The investigated structures under both pumping schemes exhibit a clear S-curve in the double log I/O plot along with drastically reduced linewidth reaching the resolution limit of around 60 µeV of our monochromator with 500 mm focal length, clearly signaturing the laser transition. By Pseudo-Voigt lineshape fitting[18] we determine an experimental Q-factor of 12200 with the dominant mode at the lasing threshold. As shown in Fig. 3(b) and 3(e), far above the lasing threshold, we observe a mode crossing behavior originating from the complex gain-competition dynamics where the dominant mode suddenly switches from mode 2 to mode 1 with further increasing pump power.[19,20]

Our micropillar lasers operate in the weak coupling regime as conventional photon lasers based on excitonic gain provided by the integrated QDs.[21] To qualitatively characterize the micropillar lasers, we fit the I/O curve by solving the laser rate equations and expressing the pump power $P_{pump}$ as a function of the output power $P_{out}$:[11,22]

$$P_{pump}(A, B, \beta, \xi') = A \frac{\gamma}{\beta} \left[ \frac{BP_{out}}{1 + BP_{out}} (1 + \xi)(1 + BP_{out}) - \beta \xi BP_{out} \right]$$

The input scaling factor $A \propto (\eta_{PCE} \delta)^{-1}$ links the pump rate to the measured pump power, where $\eta_{PCE}$ is the power conversion efficiency, $\delta$ is efficiency that converts the absorbed pump power into QDs excitons which provide optical gain. The output scaling factor $B$ factor links the intracavity photon number to the output power. $\xi = n_0 \beta / \gamma \tau_{sp}$ is a dimensionless factor including the exciton number $n_0$ at transparency threshold, the spontaneous emission factor $\beta$, the cavity decay $\gamma$, and the spontaneous emission factor $\tau_{sp}$. In order to unambiguously fit the equation, we take $\tau_{sp} = 1$ ns and $n_0 = 2.9 \times 10^3$ as have been reported for structures with nominally similar QD gain medium.[11]

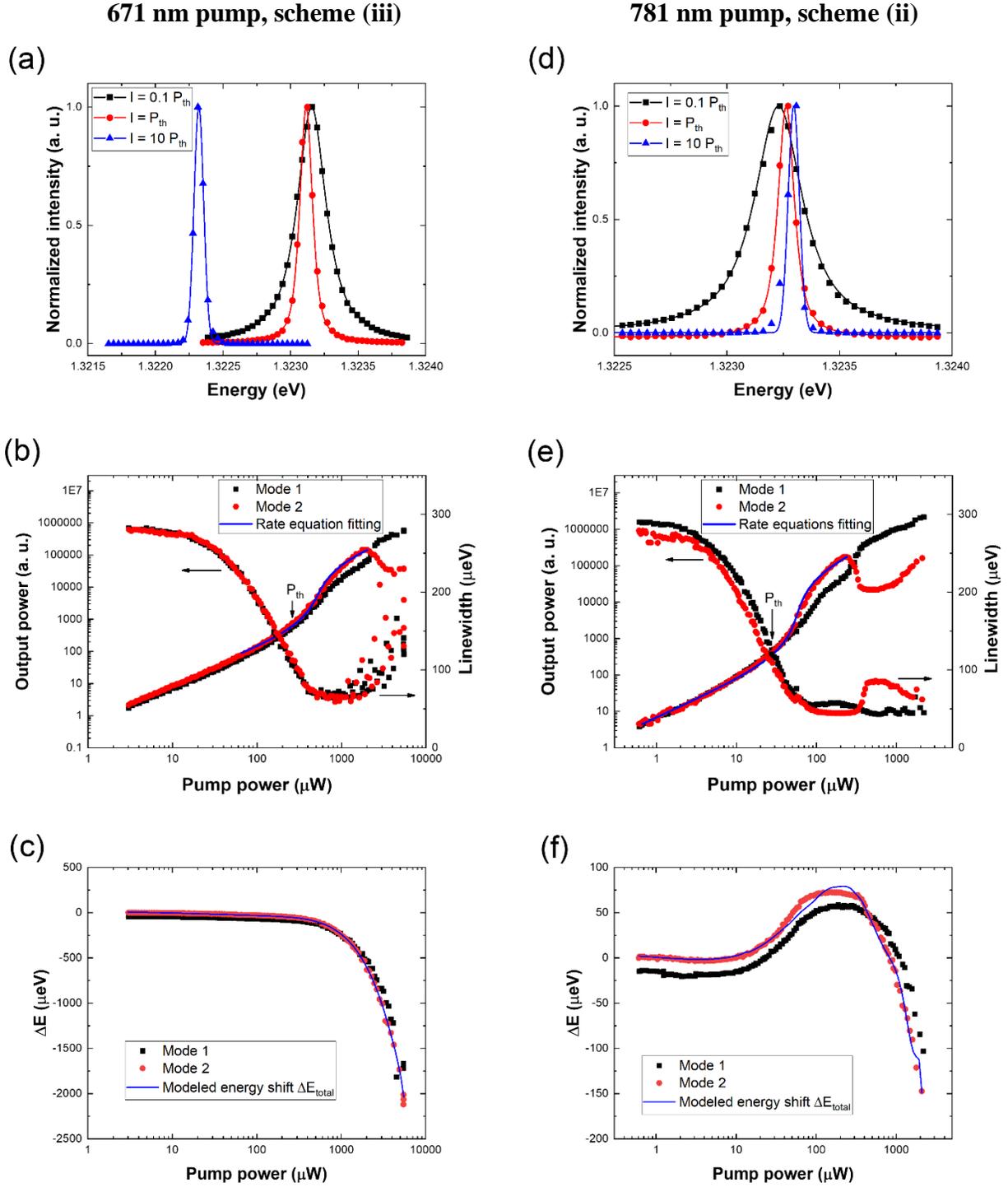

FIG. 3. I/O measurement of the micropillar laser with a diameter of 3.6 µm based on a microcavity with $Al_{0.2}Ga_{0.8}As/Al_{0.9}Ga_{0.1}As$ DBRs for two different excitation wavelength at 671 nm (left) to 781 nm (right). (a) (d) spectra of mode 2 at 0.1 $P_{th}$, $P_{th}$, and 10 $P_{th}$ (scattered dots) along with Pseudo-Voigt fitting (solid lines). (b) (e) I/O curve in double log plot and the spectral linewidth of the modes. The blue line is the fitted curve by the laser rate equations. (c) (f) The pump-power-dependent mode energy shift $\Delta E$. For a clear vision, $\Delta E$ is zeroed according to the energy of mode 2 at the lowest pump power in each plot. Black and red scattered dots are orthogonal linearly polarized fundamental modes, labeled as mode 1 and mode 2, respectively. Modeled energy shift $\Delta E_{total}$ is shown as blue line fitted to Mode 2, which takes into account the contributions of free-carrier dispersion and thermal effects.

The lasing threshold, defined at the pump power when the mean photon number in the mode is unity, can be expressed as $P_{th} = A(\xi(1-\beta)+1+\beta)\gamma/2\beta$.[22] Comparing Fig. 3(b) and 3(e), we observe a remarkable lasing threshold $P_{th}$ reduction of 9 times from (263.3 ± 14.0) μW to (28.8 ± 1.6) μW by simply changing the pump laser wavelength from 671 nm to 781 nm and therefore entering the low-absorbing pumping scheme (ii) of the microcavity. The observed lasing threshold reduction factor is nearly identical to the 10 times PCE change as estimated in Fig. 1(b), demonstrating that the reduction of pump laser absorption in the top DBRs is a leading factor to improving the pump efficiency.

It is interesting to note that optical absorption in the top DBRs of conventional microcavities also leads to unwanted heating of the devices, which can be monitored by observing the redshift of the mode emission energy.[23] Undesired heating not only obstructs the realization of a spectrally homogeneous micropillar laser array for neuromorphic computing, which requires a inhomogeneous broadening less than 200 μeV,[5,9] but also hinders the study of physical effects such as high-β lasing in such microstructures. In the absorbing pumping scheme Fig. 3(c), both modes redshift monotonously due to a strong thermal effect; while in the absorption-free scheme Fig. 3(f), we could observe obvious effects from free-carrier dispersion, namely carrier-induced change of the refractive index, before the thermal effects dominate. Under low pump power, the modes first exhibit a minor redshift of around 4 μeV resulting from bandgap shrinkage, and then a major blueshift around 70 μeV that can be attributed to band filling and the plasma effect.[23-25] Under high pump power, despite the blueshift coming from increasing free carrier density, thermal effects dominate and result in a strong redshift. The high-absorbing case exhibits a redshift from $0.1P_{th}$ to $10\,P_{th}$ as high as (0.83 ± 0.13) meV, while the low-absorbing case exhibits a small blueshift of only (71.3 ± 1.8) μeV. Note that all effects do coexist in both pumping schemes, while in the low-absorbing pumping scheme the plasma effect is more dominant thanks to the weakened thermal effects. With the related free-carrier dispersion data from H. C. Huang *et al.*,[25] we present a simple model as well to show the contributions as a function of optical pump power taking into accounts the above-mentioned effects (see the supplementary material).

Additionally, in the absorbing scheme, the emission modes above lasing threshold become unstable and fluctuate due to thermal effects, leading to an inhomogeneous broadening of the laser linewidth as we can see in Fig. 3(b). In the absorption-free scheme, the modes, however, remain stable over the whole pump power range. This even allowed us to observe the linewidth broadening of the weak mode when the other stronger mode dominates and takes over lasing, highlighting an often-overlooked characteristics that one can only see in the absence of severe heating.

Further pump-wavelength dependent I/O measurements are performed on the fabricated-micropillar lasers with diameters of 2.9 μm, 3.6 μm, and 5.4 μm for a broader pump laser wavelength range, covering all three pumping schemes to obtain comprehensive insight into the lasing characteristics. Again, the lasing thresholds are extracted via laser rate equations and are shown in Fig. 4(a). The Q-factors extracted at lasing threshold are 12000, 12200, and 15000, respectively. As expected, under the wetting layer pumping scheme (iii) where the laser wavelength is longer than 820 nm, the micropillar lasers exhibit a high lasing threshold above several hundreds of μW at 833 nm and increases even up to mW scale at 899 nm due to a weaker pump absorption by the wetting layer. Under the low-absorbing pumping scheme (ii) between 700 nm to 820 nm, the lasing threshold drastically reduces to tens of μW and even reaching a record low value of $(12.8 \pm 0.3)$ μW pumped at 708 nm for the 2.9 μm micropillar laser (see the supplementary material). Finally, when the laser wavelength is shorter than 700 nm in absorbing scheme (iii), the lasing threshold increases rapidly again up to several hundreds of μW under 671 nm pump laser. Additionally, although not included in the plot, the micropillars could not complete the laser transition when pumped by a 532 nm laser due to severe heating of the devices. The lasing thresholds depicted in Fig. 4(a) clearly signature an inverse bell-shape as predicted by the simulation of PCE.

We also obverse that the 2.9 μm pillar in general shows a lower lasing threshold than the 5.4 μm pillar. This can be attributed to smaller mode volume and therefore higher Purcell- and $\beta$-factor of the smaller micropillar laser. The averaged extracted $\beta$-factors are $(3.8 \pm 4.6)$%, $(2.3 \pm 1.2)$%, and $(1.0 \pm 1.3)$% for the pillars with 2.9 μm, 3.6 μm, and 5.4 μm diameter, respectively. Meanwhile, from the inverse of the input scaling factor $A$ in the rate equations fit, we obtain the pump efficiency plotted in Fig. 4(b), which is a product of PCE and the exciton conversion efficiency $\delta$. We could clearly observe an increase of pump efficiency in pumping scheme (ii). Note that due to the complex mode competition and crossing behavior above the lasing threshold, the extraction of the $\beta$-factor and pump efficiency becomes less accurate. Interestingly, despite a lower lasing threshold from the smaller pillar, its pump efficiency is lower than for the large diameter pillar. This behavior can be attributed to the finite pump laser spot size, which is bigger than the fundamental mode area of the small pillars, leading to a reduced fundamental mode pump efficiency (see the supplementary material).[26]

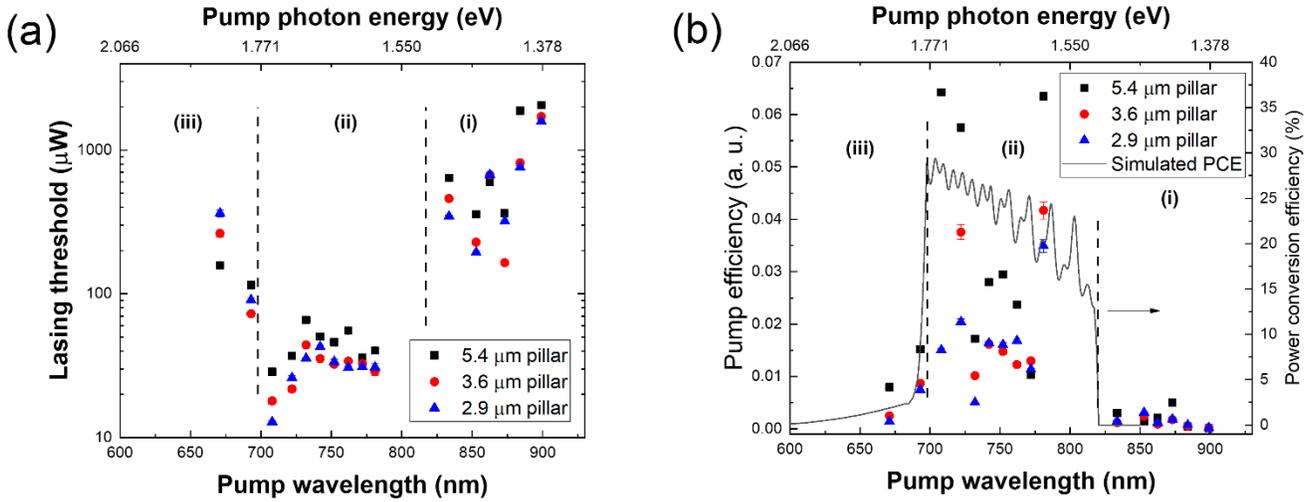

FIG. 4. Extraction of the (a) lasing threshold and the (b) pump efficiency with the laser rate equation fitting. Three pumping scheme (i-iii) are marked in the plots, where a huge difference of the lasing threshold and the pump efficiency can be readily seen.

In conclusion, we demonstrated that a modification of the epitaxial layer design of planar QD-microcavity structures, namely the change of binary GaAs in the DBRs by ternary $Al_{0.2}Ga_{0.8}As$, leads to a low-absorbing pumping scheme. The reduction in threshold pump power of micropillar lasers can easily reach more than an order of magnitude, from $(363.0 \pm 18.5)$ µW pumped at 671 nm to values as low as $(12.8 \pm 0.3)$ µW pumped at 708 nm. Simulations predict that the upper limit of the associated PCE can be increased by nearly 6 times under ideal pumping comparing the conventional design to our modified design. The reduced lasing threshold strongly enables scaling up the number of micropillar lasers in applications requiring high energy efficiency. It also reduces undesired thermal heating effects, leading to more stable emission characteristics. The developed design is therefore highly interesting for, both, the fundamental study and applications of micropillar devices.

Note that the design of the devices and the measurements in this study are all conducted under 77 K to benefit from sufficiently high optical gain with a single layer of QDs. To provide sufficient optical for application relevant room temperature operation one can increase the carrier confinement by including AlGaAs barriers near the QD layer or by introducing stacked layers of QDs,[27] as well as using optimized bottom DBR designs to further increase the PCE. For instance, one could additionally modify the bottom DBRs design so that it would also be reflecting at the pump laser wavelength.[28] On the other hand, to extend the low-absorbing pump window, one could further increase the Al composition in the AlGaAs alloy at the cost of a reduced refractive index contrast and therefore more required pairs of DBRs. Additionally, alternative to an epitaxially grown upper DBR, one could combine large bandgap dielectric DBRs deposited by means of plasma electric chemical vapor deposition (PECVD).[29] Our work provides a crucial insight into the fundamental design of micropillar lasers and paves the way to scaling up the number of photonic neurons in neuromorphic schemes based on coupled laser array.

See the supplementary material for the additional information on I/O plots, optical numerical simulations, TMM simulations of reflectance and transmittance, modeling of the mode energy shift resulted from thermal effects and free-carrier dispersion, as well as diameter-dependent fundamental mode pump efficiency.

## Acknowledgement

This work is funded by German Research Foundation (Re2974/20-1, Re2974/21-1, Re2974/26-1, INST 131/795-1 FUGG) and Volkswagen Foundation (NeuroQNet II). We would like to thank K. Schatke, R. Schmidt, and R. Linke for the technical support on epitaxy and processing. We would like to thank J. Große, T. Heuser, and S. Richter for the assistance and discussion.

## Author Declarations

### Conflict of Interest

The authors have no conflicts to disclose.

### Author Contributions

**Ching-Wen Shih**: conceptualization (lead), data curation (lead), formal analysis (lead), software (lead), visualization (lead), resources (supporting), investigation (lead), methodology (equal), writing/original draft preparation (lead). **Imad Limame:** methodology (equal), resources (lead), investigation (support). **Sebastian Krüger:** formal analysis (support), investigation (support). **Chirag Palekar:** methodology (support), investigation (support). **Aris Koulas-Simos**: methodology (support), investigation (support). **Daniel Brunner**: funding Acquisition (equal), validation (equal). **Stephan Reitzenstein:** funding Acquisition (equal), supervision (lead), validation (equal), writing/review & editing (lead).

### Data Availability

The data that support the findings of this study are available from the corresponding author upon reasonable request.

# Supplementary material: Low-threshold lasing of optically pumped micropillar lasers with $Al_{0.2}Ga_{0.8}As/Al_{0.9}Ga_{0.1}As$ distributed Bragg reflectors

Ching-Wen Shih,[1] Imad Limame,[1] Sebastian Krüger,[1] Chirag C. Palekar,[1] Aris Koulas-Simos,[1] Daniel Brunner,[2] Stephan Reitzenstein[1]

[1]*Institut für Festkörperphysik, Technische Universität Berlin, Hardenbergstraße 36, D-10623 Berlin, Germany*

[2]*Département d'Optique P. M. Duffieux, Institut FEMTO-ST, Université Franche-Comté CNRS UMR 6174, Besançon, France*

1. **Transmittance and reflectance of the microcavity with $Al_{0.2}Ga_{0.8}As/Al_{0.9}Ga_{0.1}As$ DBRs design**

In addition to the PCE shown in the main text, Fig. S1 depicts the TMM simulation reflectance and transmittance of the design with modified DBRs. The oscillation in both the reflectance and transmittance lead to the oscillation behavior in PCE. Meanwhile, despite the improved PCE, around 30% of the pump power is still reflected while the other around 40% of the pump power is transmitted to the GaAs substrate.

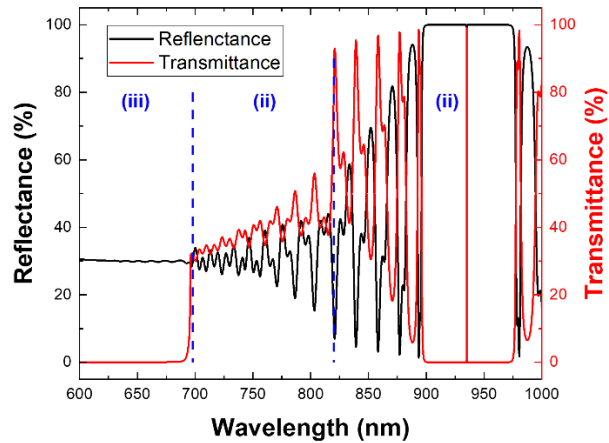

FIG. S1. Reflectance (black) and transmittance (red) of the microcavity simulation by TMM.

## 2. Optical numerical simulation

Results of optical numerical simulations performed via the finite element method (FEM) by JCMsuite are presented in Fig. S2 to show the Q-factor, mode volume, and Purcell-factor. Both, the above-mentioned modified $Al_{0.9}Ga_{0.1}As/Al_{0.2}Ga_{0.8}As$ DBRs design and conventional $Al_{0.9}Ga_{0.1}As/GaAs$ DBRs, are simulated with a fixed micropillar diameter of 5 µm. To ensure directional surface emission, there are always 3 less mirror pairs in the top DBR than in the bottom DBR. As we can see in Fig. S2(a) and S2(b), the Q-factor and mode volume of the modified DBRs optical cavity are lower than those of the one with conventional DBRs due to the slight decrease of the refractive index contrast in the modified DBR design as shown in the upper panels of Fig. 2 in the main text. As a result, the Purcell factor as shown in Fig. S2(c), which scales with $\frac{Q}{V}$, also appears to be lower in the modified microcavity design. Nevertheless, the decreased Q-factor, and therefore Purcell factor, could be easily mitigated by increasing the number of mirror pairs, leading to higher reflectivity while maintaining almost the same mode volume. In fact, additional 6~7 mirror pairs in both the bottom and top DBR are enough to maintain the same micropillar cavity performance.

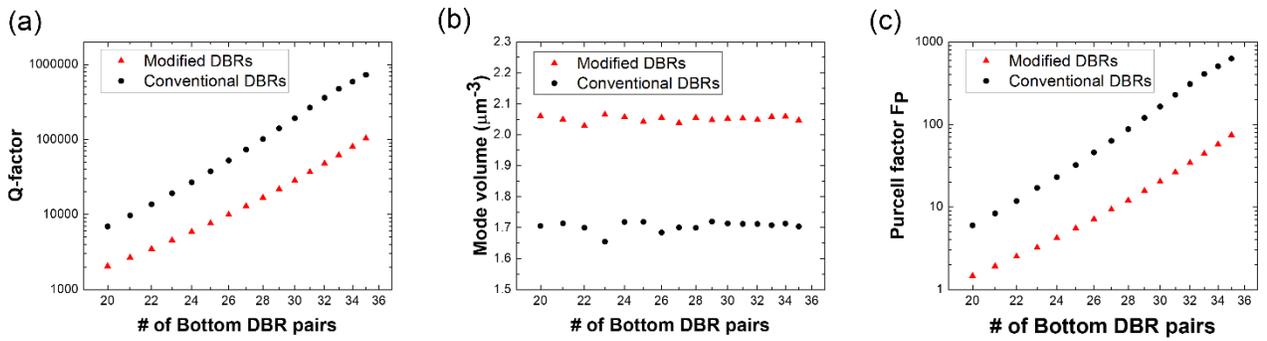

FIG. S2. (a) Q-factor, (b) mode volume, and (c) Purcell factor vs. the number of DBR mirror pairs for modified and conventional case.

## 3. Carrier- and thermal induced mode energy shift

Both free-carrier dispersion effects and thermal effects induced by the optical pump laser could lead to shifts in the mode emission peak ΔE. Depending on the strength of each effect, ΔE could be distinctive as shown in Fig. 3(c) and Fig. 3(f) in the main text. In order to distinguish the contributions from both effects, we first analyze the ΔE under low-absorbing pumping scheme (ii) with the 781 nm pump laser. From Fig. S3(a) we could immediately observe that under high pump power the smaller pillar experiences a larger redshift due to their worse heat dissipation ability. However, the behavior differs at low pump power, where the effect of heating is not prominent and the free-carrier dispersion effects dominate.

With the carrier induced refractive index change of the GaAs cavity $\Delta n_{GaAs}$ from Fig. 6 in [30], we simulated the mode energy shift as a function of $\Delta n_{GaAs}$ and thus as a function of the carrier concentration N, in the GaAs cavity by means of transfer matrix method as shown in Fig. S3(b). We further make a simple assumption that the absorbed pump power within each layer, which is a product of the layer's PCE and the pump laser power $P_{pump}$, is proportional to the created carrier concentration N in the corresponding layer $N = P_{pump} \times PCE \times C / d$, where d is the layer thickness and C is a fitting parameter. We therefore obtain $\Delta E_{carrier}(P_{pump}, N)$ and could fit to the experimental mode energy shift under low excitation power as shown in the black solid curve in Fig. S3(c). Finally, from the difference between the experimental $\Delta E$ and calculated $\Delta E_{carrier}$, we fit the remaining thermal contribution $\Delta E_{thermal}$, which is accountable for the redshift, with a 6$^{th}$ order polynomial function as listed in Table S1. As such, we show the contributions from both effects in Fig S3(c) for the 3.6-μm-diameter pillar. Figure S3(d) further shows the mode shifting contributions under the absorbing scheme (iii) with the 671 nm pump laser. Additionally, in order to achieve a comparable estimation, the same fitting parameter $C = 2.218 \times 10^{18}$ cm$^{-2}$W$^{-1}$ is used for all cases to calculate the refractive index change in all absorbing layers as a function of laser pump power. With the developed model, we also fit to the micropillars with different diameter as shown in the solid lines in Fig. S1(a) while assuming the fitting parameter C is diameter independent, which in other words, assuming the free-carrier contribution is the same for all three micropillars. The extracted thermal contributions are shown in the subplot of Fig. S1(a), where we can observe a stronger thermal contribution in the smaller micropillars.

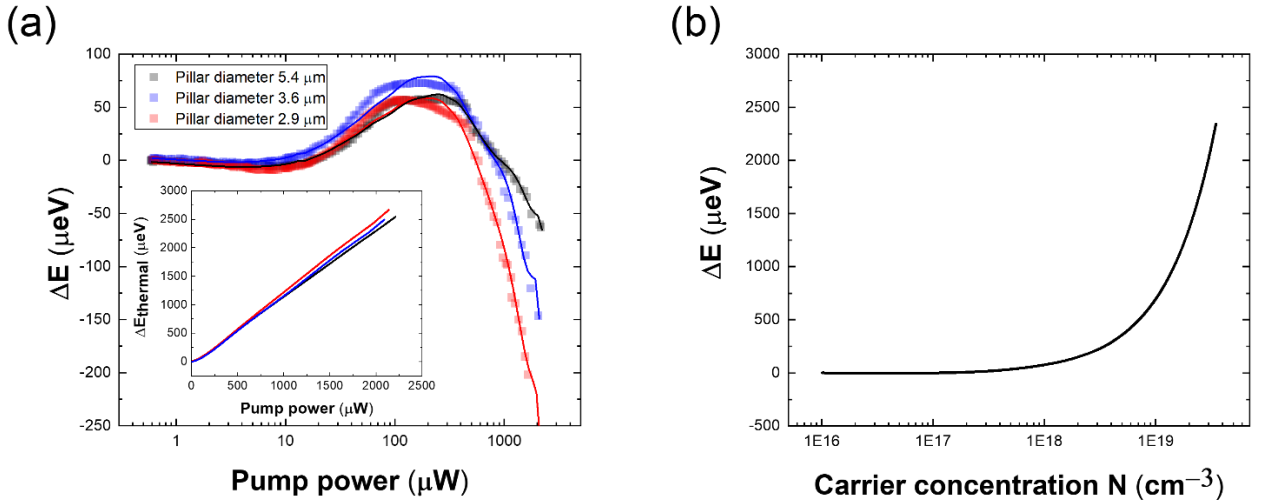

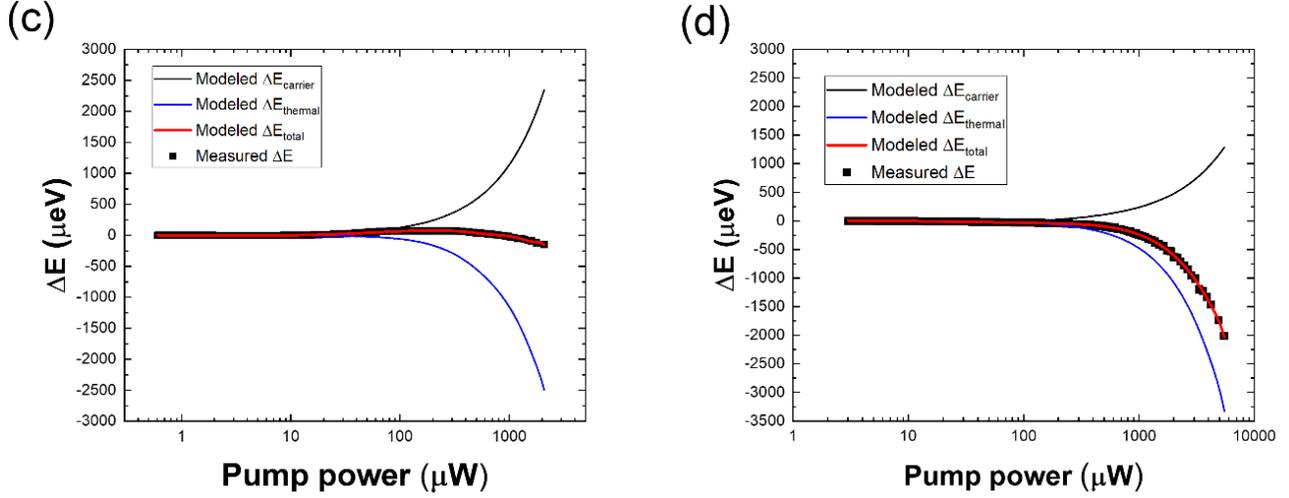

FIG. S3. (a) Mode energy shift as a function of pump power for different diameter pillars under low-absorbing pumping scheme. Scattered dots are the experimental data; solid lines are the modeled results. Subplot: the mode energy shift from thermal contribution. (b) The simulated mode energy shift as a function of the carrier concentration N in the GaAs cavity resulted from free-carrier dispersion under the low-absorbing pumping scheme. (c) and (d) show the modeled and experimental mode energy shift of the 3.6 µm-diameter pillar under 781 nm and 671 nm pump laser, respectively. Black solid curve: the modeled carrier contribution to the mode energy shift $\Delta E_{carrier}$; blue solid curve: the modeled thermal contribution to the mode energy shift $\Delta E_{thermal}$; red curve: the total modeled mode energy shift $\Delta E_{total}$; Scattered dots: experimental mode energy shift.

|  | $A_0$ [µeV] | $A_1$ [µeV/µW] | $A_2$ [µeV/µW$^2$] | $A_3$ [µeV/µW$^3$] | $A_4$ [µeV/µW$^4$] | $A_5$ [µeV/µW$^5$] | $A_6$ [µeV/µW$^6$] |
|---|---|---|---|---|---|---|---|
| 781 nm pump laser, 5.4-µm micropillar | -1.94 | -0.58 | -2.12E-3 | 3.17E-06 | -2.31E-09 | 8.14E-13 | -1.11E-16 |
| 781 nm pump laser, 3.6-µm micropillar | 1.50 | -0.31 | -3.56E-3 | 6.06E-06 | -5.05E-09 | 2.02E-12 | -3.10E-16 |
| 781 nm pump laser, 2.9-µm micropillar | -3.10 | -0.50 | -2.75E-3 | 4.36E-06 | -3.48E-09 | 1.36E-12 | -2.06E-16 |
| 671 nm 3.6-µm micropillar | 4.08 | -0.33 | -2.39E-4 | 1.18E-07 | -3.59E-11 | 5.84E-15 | -3.78E-19 |

TABLE S1. The 6$^{th}$ order polynomial function to fit the thermal contributions. $\Delta E_{thermal}(P_{pump}) = \sum_{i=0}^{6} A_i P_{pump}^i$.

From Fig. S3 we could see that there is always a strong competition and intermix of the free-carrier dispersion effects and the thermal effects. Note that although it seems to be contradicting that under the same pump laser power, the low-absorbing pumping case has larger thermal contribution than the other case, this can be in fact be attributed to an inhomogeneous temperature gradient in the microcavity. For the low-absorbing case, the pump laser is absorbed and results in heating of the central GaAs cavity directly, which has a high impact on the emission mode energy. However, in the other case, the pump laser is mostly absorbed by the top few DBRs, which causes the refractive index change in the top layers far away from the GaAs cavity, leading to less direct impact on the emission mode energy. Nevertheless, the low-absorbing pumping scheme has a strongly reduced lasing threshold and the two cases should not be compared under the same pump power. On the other hand, the assumption that the carrier concentration scales linearly with the laser pump power can deviate from reality as the lasing mechanism could clamp the carrier concentration in the active area, leading to an over-estimated free-carrier dispersion effects and therefore thermal effects in the model.

### 4. Diameter-dependent pump efficiency

Figure 2(b) in the main text shows the calculated PCE considering the microcavity along the pillar axis. However, in reality, the pillar diameter has to be taken into accounts as well when the focused pump laser spot size is finite. From Fig. 4 in the main text, we observe that the smaller pillars exhibit a lower pump efficiency despite their lower threshold due to the spatial mismatch of the pump laser spot to the fundamental mode area. By means of FEM simulations, we calculated the fundamental mode area, defined as the FWHM of the in-plane field intensity, to be 5.0 $\mu m^2$, 2.4 $\mu m^2$, 1.5 $\mu m^2$, for the 5.4-$\mu$m, 3.6-$\mu$m, and 2.9-$\mu$m pillars, respectively.

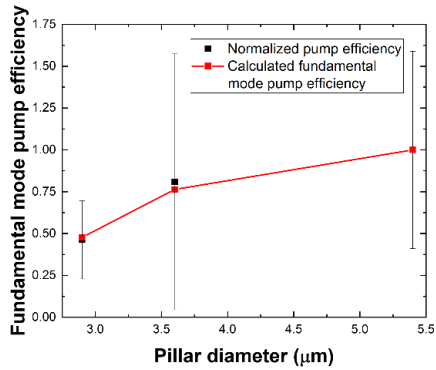

FIG. S4. Normalized fundamental mode pump efficiency of the 5.4-**$\mu$**m, 3.6-**$\mu$**m, and 2.9-**$\mu$**m micropillars. Black scattered dots: experimental pump efficiency averaged over the low-absorbing scheme. Red scattered data: calculated fundamental pump efficiency.

We further make a simple assumption that only the 2-$\mu$m-diameter pump laser spot within this fundamental mode area can contribute to the excitation and therefore calculate the percentage of power conversion into the fundamental mode. We compare the calculated efficiencies to the experimental efficiencies averaged over the low-absorbing scheme and normalize them to the 5.4-$\mu$m pillar. The estimated diameter-dependent pump power efficiency agrees very well with the measured results as shown in Fig. S4, where a reduction in the pump efficiency with smaller pillars is predicted and observed, proving that the pump efficiency is strongly influenced by the overlap between the fundamental mode area and the pump laser spot. The lasing threshold on the other hand, is also influenced by the Purcell factor and β-factor, where the smaller pillar takes lead and in the end results in a lower lasing threshold.

5. **Additional I/O plots**

Figure S5 shows the I/O measurement of the 2.9-μm-diameter micropillar with $Al_{0.2}Ga_{0.8}As/Al_{0.9}Ga_{0.1}As$ DBRs pumped at 708 nm. A lowest lasing threshold $P_{th} = (12.8 \pm 0.3)$ μW is extracted by the laser rate equation fitting.

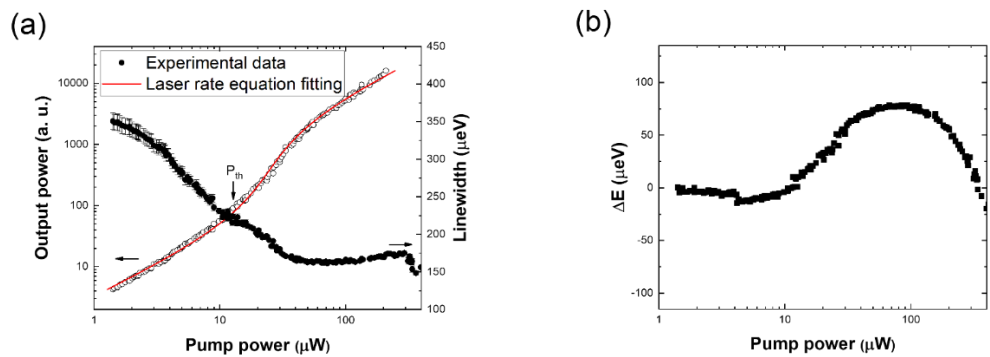

FIG. S5. I/O measurement of the dominant laser mode based on a microcavity. (a) Output power and laser emission linewidth as a function of the pump power. Here, the $\beta$-factor is fitted to be $(13.8 \pm 0.4)\%$. (b) The pump-power-dependent mode energy shift ΔE.